\documentstyle[epsfig]{qcdparis}
\def\lapproxeq{\lower .7ex\hbox{$\;\stackrel{\textstyle <}{\sim}\;$}}
\def\gapproxeq{\lower .7ex\hbox{$\;\stackrel{\textstyle >}{\sim}\;$}}

\begin{document}

\title{\hfill DTP/95/60\eject Description of $F_2$ and the gluon at small $x$}

\author{A.\ D.\ Martin}

\affil{Department of Physics, University of Durham, DH1 3LE, UK.}

\abstract{We give a brief overview of the perturbative QCD
description of the proton deep-inelastic structure function $F_2
(x, Q^2)$ at small $x$.  We discuss GLAP and BFKL approaches, and
then we review progress towards a more unified treatment.}

\resume{Nous d\'{e}crivons bri\`{e}vement la fonction de
structure $F_2 (x, Q^2)$ du proton aux petites valeurs de $x$,
dans l'approche perturbative de la chromodynamique quantique.
Nous discutons les approches GLAP et BFKL, et passons en revue
les progr\`{e}s r\'{e}cents vers un traitement plus unifi\'{e}.}

\twocolumn[\maketitle]
\fnm{7}{Contribution to the Workshop on Deep Inelastic Scattering
and QCD, Paris, April 1995}

\section{Introduction}
Fig.\ 1 is a sketch of the gluon content of the proton.  GLAP or
Altarelli-Parisi evolution to higher $Q^2$ increases the
resolution $1/Q$ in the transverse plane, while BFKL evolution to
small $x$ builds up to gluon density until gluon recombination,
or shadowing, can no longer be neglected.  At small $x$ the
behaviour of $F_2 (x, Q^2)$ follows that of the gluon on account
of the $g \rightarrow q\overline{q}$ transition.  First we
briefly discuss the GLAP and BFKL descriptions of the
measurements of $F_2 (x, Q^2)$ at HERA, and then we review the
progress made towards a unified treatment which incorporates both
GLAP and BFKL dynamics.
\ffig{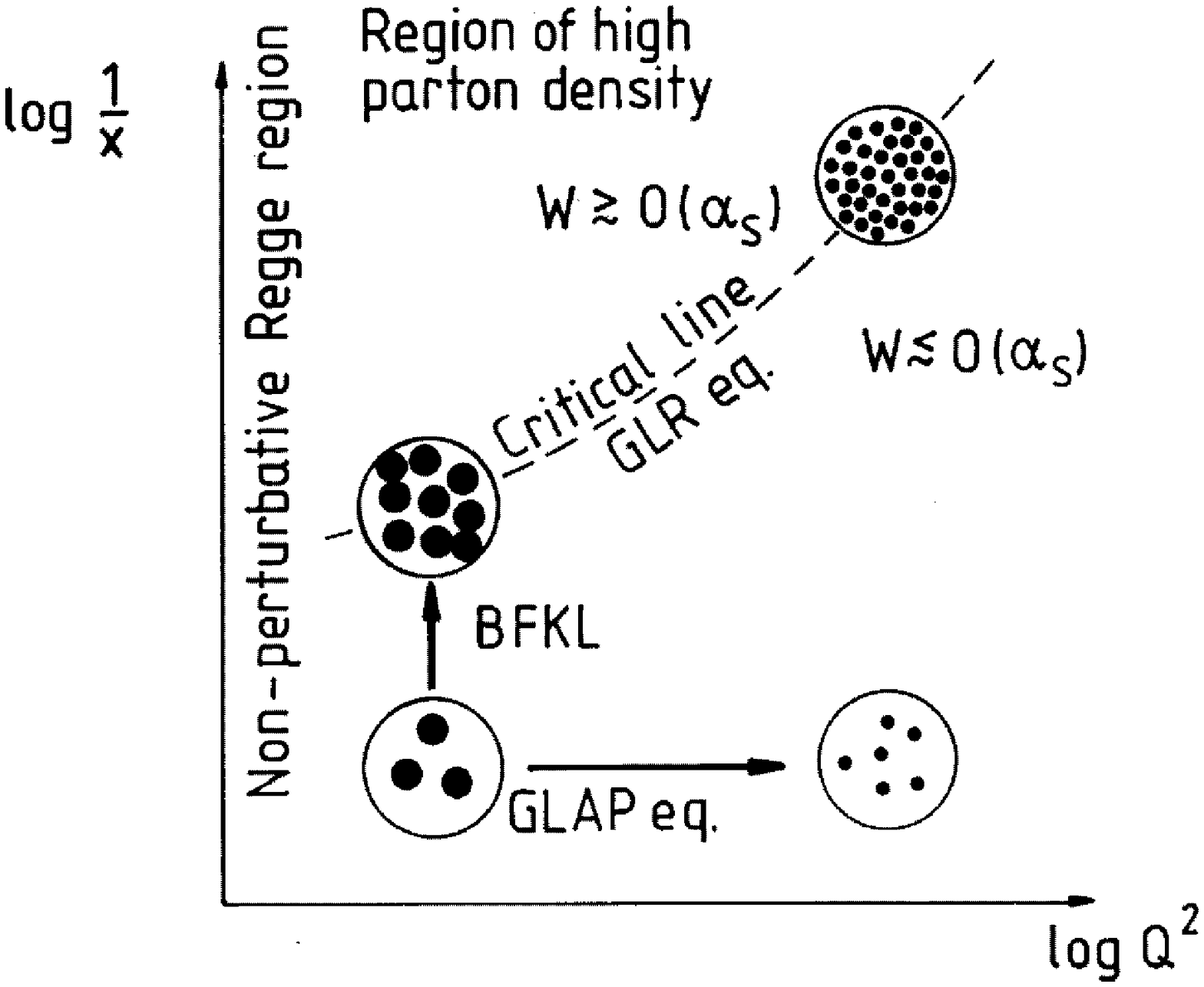}{60mm}{\em The gluonic content of the proton as
\lq\lq seen" in different deep inelastic $(x,Q^2)$
regions.  The critical line, where gluon recombination becomes
significant, occurs when $W \approx 0(\alpha_S)$.  $W$ is the
ratio of the quadratic to the linear term on the right hand side
of equation (9).}{fig1}
\section{GLAP description}
%
%
There have been several successful attempts to describe the HERA
data \cite{h1,zeus} based on GLAP evolution
\cite{mrsg,cteq,grv,bf}.  Fig.\ 2, which shows $F_2$ at $x = 4
\times 10^{-4}$, illustrates some of the main features.  At $Q^2
= 4$ GeV$^2$, say, we have
\begin{eqnarray}
F_2 & \sim & x S \sim A_S \: x^{- \lambda_S} \label{a1} \\
\frac{\partial F_2}{\partial \log Q^2} & \sim & xg \sim A_g \:
x^{- \lambda_g}. \label{a2}
\end{eqnarray}
\noindent That is, to a good approximation, $F_2$ determines the
sea quark $(S)$ distribution while the slope determines the gluon
$(g)$.  The MRS(A$^\prime$) and MRS(G) curves correspond to two
global fits to deep-inelastic and related data.  The former has
$\lambda_g \equiv \lambda_S \simeq 0.2$, while in the latter both
$\lambda_g$ and $\lambda_S$ are taken as free parameters with
$\lambda_g \simeq 0.35$ and $\lambda_S \simeq 0.1$, that is a
\lq\lq steep" gluon and a \lq\lq flat" sea distribution.

The GRV description \cite{grv} is obtained by evolving from
valence-like parton distributions at a very low scale $Q_0^2 =
0.3$ GeV$^2$.  By $Q^2 = 4$ GeV$^2$ the small $x$ behaviour of
$xg$ and $xS$ approximates the double leading logarithmic (DLL)
form
\begin{equation}
\exp \: \left ( 2 \left [ \frac{36}{25} \log \left ( \log
\frac{Q^2}{\Lambda^2} / \log \frac{Q_0^2}{\Lambda^2} \right ) \:
\log \left ( \frac{1}{x} \right ) \right ]^{\frac{1}{2}} \right
).
\label{a3}
\end{equation}
\noindent Although not as \lq\lq steep" as $x^{- \lambda}$, it
can be approximated by this form over a limited interval of $x$.
The effective value for GRV partons is $\lambda_g \simeq
\lambda_S \simeq 0.28$, see Fig.\ 2.  Ball and Forte \cite{bf}
also find that the HERA $F_2$ data can be well described by DLL
forms.  In the DLL approach the effective slope $\lambda$ can be
decreased (increased) by simply increasing (decreasing) $Q_0^2$.
\ffig{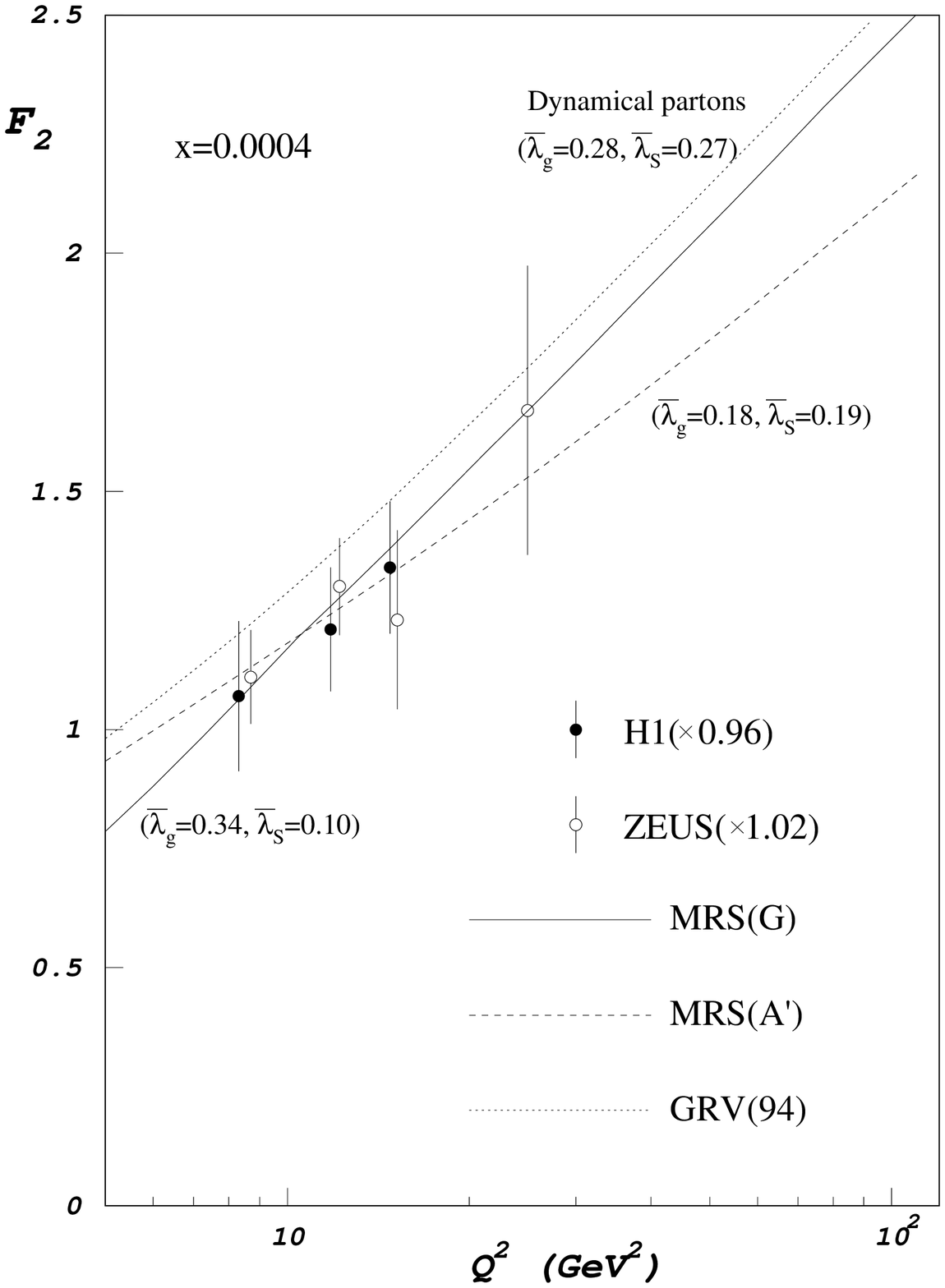}{100mm}{\em HERA data [1, 2] for
$F_2^{ep}$ compared with MRS(A$^\prime$,G) [3] and GRV
[5] partons.  The effective values of $\overline{\lambda}$,
obtained from $xg \sim x^{- \overline{\lambda}_g}$ and $x S \sim
x^{- \overline{\lambda}_S}$, are shown.  The figure is taken from
ref.\ [3].}{fig2}
\section{BFKL description}
The BFKL description of $F_2$ at small $x$ is based on the
$k_T$-factorization formula \cite{cch}
\begin{equation}
F_2 (x, Q^2) = \int_x^1 \frac{dx^\prime}{x^\prime} \int
\frac{dk_T^2}{k_T^2} \: f (x^\prime, k_T^2) \: \hat{F}_2 \left (
\frac{x}{x^\prime}, \frac{k_T^2}{Q^2}, \alpha_S \right )
\label{a4}
\end{equation}
\noindent where $f (x^\prime, k_T^2)$ is the gluon distribution
{\it unintegrated} over $k_T$, while $\hat{F}_2$ is the structure
function of a gluon of virtuality $k_T^2$ probed by a photon of
virtuality $Q^2$, that is the contribution of the subprocess
$\gamma g \rightarrow q\overline{q}$.  $f$ is calculated by
integrating the BFKL equation
\begin{equation}
- x \: \partial f/\partial x = K \otimes f
\label{a5}
\end{equation}
\noindent down in $x$ from a starting distribution at $x
= x_0 = 0.01$, say.  In this symbolic form of the equation
$\otimes$ represents a convolution over $k_T$.  The BFKL equation
\cite{bfkl} effectively sums the
leading $\alpha_S \log (1/x)$ contributions
\begin{equation}
f \sim \exp (\lambda \log (1/x)) \sim x^{- \lambda}
\label{a6}
\end{equation}
\noindent where $\lambda$ represents the largest eigenvalue of
the BFKL kernel $K$; $\lambda = 12 \alpha_S \log 2/\pi$ for fixed
$\alpha_S$ \cite{bfkl} and $\lambda \simeq 0.5$ for running
$\alpha_S$ \cite{akms}.

The solution of BFKL equation is sensitive to the treatment of
the infrared (non-perturbative) region.  For running $\alpha_S$
it is found that
\begin{equation}
f \sim C(k_T^2) x^{- \lambda}
\label{a7}
\end{equation}
\noindent where $\lambda \approx 0.5$ has much less infrared
sensitivity than the normalization $C$.  The prediction for $F_2$
follows from the $k_T$-factorization formula (\ref{a4})
\begin{eqnarray}
F_2 & = & f \otimes \hat{F}_2 + F_2^{bg} \nonumber \\
& \simeq & C^\prime (Q^2) x^{- \lambda} + F_2^{bg}
\label{a8}
\end{eqnarray}
\noindent where $\lambda \simeq 0.5$, and $F_2^{bg}$ is
determined from the large $x$ behaviour of $F_2$.  Once the
overall normalization of the BFKL term is adjusted by a suitable
choice of the infrared parameters a satisfactory description of
the $F_2$ HERA data is obtained \cite{akms}.  Indeed the
BFKL-based treatment gives a similar description to GLAP.  With
GLAP, the observed steepness is either incorporated (as a factor
$x^{- \lambda}$) in the starting distributions or generated by
evolution from a low scale $Q_0^2$.  The steepness can be
adjusted to agree with the data by varying $\lambda$ or $Q_0^2$.
On the other hand the leading $\log (1/x)$ BFKL prediction for
the shape $F_2 - F_2^{bg} \sim x^{- \lambda}$ with $\lambda
\simeq 0.5$ is prescribed.  It remains to see how well it
survives a full treatment of sub-leading effects.

Shadowing is yet another possible ambiguity.  The $x^{- \lambda}$
growth of the gluon cannot go on indefinitely with decreasing
$x$.  It would violate unitarity.  The growth is eventually
suppressed by gluon recombination, which is represented by an
additional quadratic term so that (\ref{a5}) has the form
\begin{equation}
-x \: \partial f/\partial x = K \otimes f - V \otimes f^2
\label{a9}
\end{equation}
\noindent where $V$ contains a factor $\alpha_S^2/k_T^2 R^2$.
The factor $1/R^2$ is expected; the smaller the transverse area
$(\pi R^2)$ in which the gluons are concentrated within the
proton, the stronger the effect of recombination.  If the gluons
are spread uniformly throughout the proton $(R \sim 5$ GeV$^{-
1}$) shadowing effects are predicted to be small in the HERA
regime \cite{akms}.  On the other hand if gluons are concentrated
in \lq\lq hot-spots" with, say, $R = 2 \: {\rm GeV}^{-1}$ then
shadowing gives an observable reduction in the prediction for
$F_2$, particularly at low $Q^2$ \cite{akms}.  In fact such a
description is in line with the HERA data but, of course, other
explanations are equally plausible.

In the remaining two sections we discuss the progress that is
being made towards a unified description which incorporates both
BFKL and GLAP evolution.
\section{Perturbation series for $\gamma (\omega, \alpha_S)$}
The possible onset of BFKL behaviour in the HERA small $x$ domain
has prompted several studies \cite{ekl,ehw,bf2,frt} of the
validity of GLAP evolution in this region.  The situation is well
summarised in Fig.\ 3 which shows the terms that occur in the
expansion of the anomalous dimensions as power series in
$\alpha_S$ and the moment index $\omega$.  Consider, for
simplicity, GLAP evolution for the gluon alone
\begin{equation}
\frac{\partial g (x, Q^2)}{\partial \log Q^2} = \int_x^1
\frac{dy}{y} \: P_{gg} \left ( \frac{x}{y} \right ) g (y, Q^2).
\label{a10}
\end{equation}
\ffig{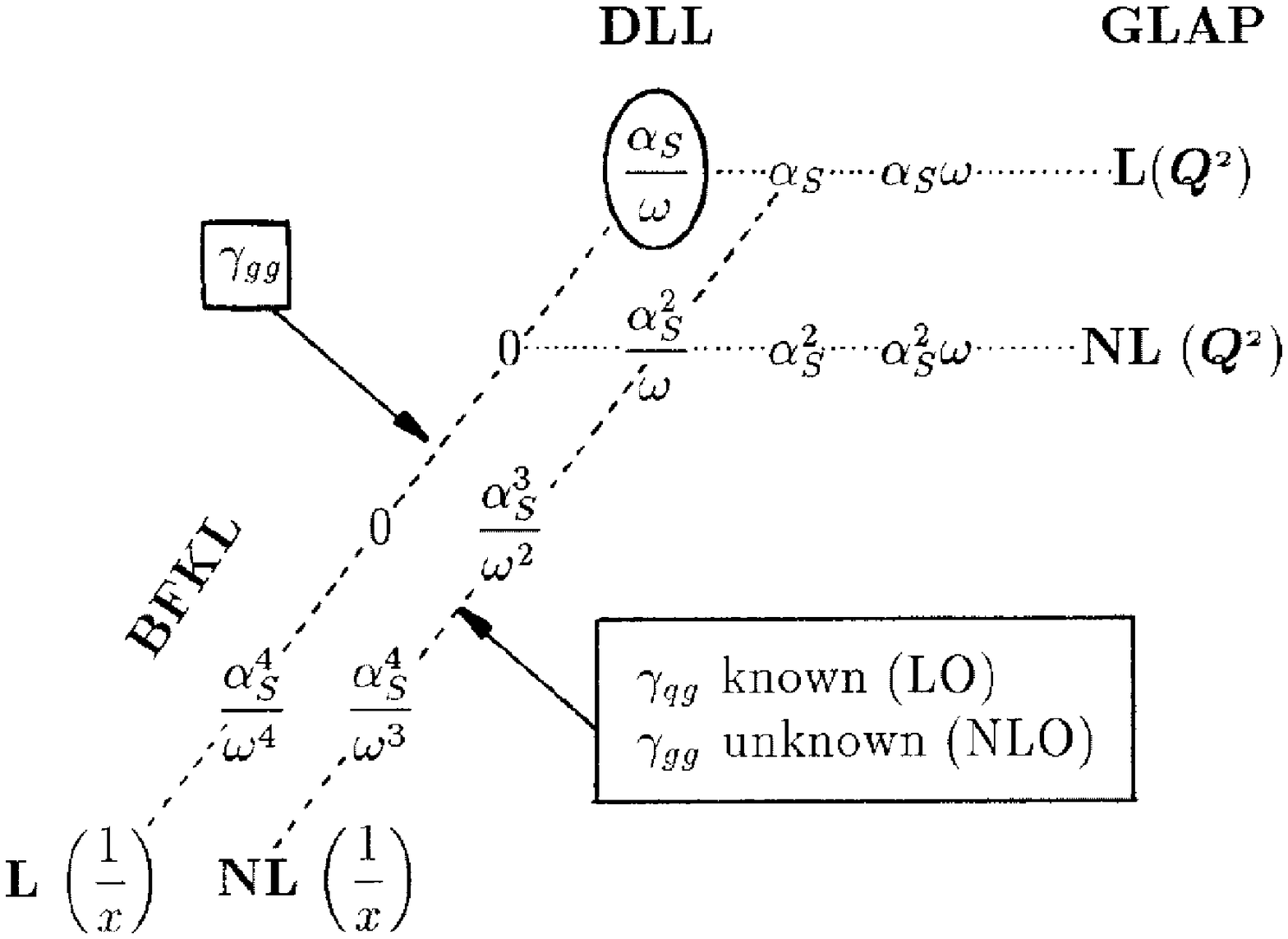}{65mm}{\em Possible terms in the perturbative
expansion of the anomalous dimensions (and associated splitting
functions).  Leading order GLAP and BFKL have only the DLL term
in common.}{fig3}
\noindent In moment space this takes the factorized form
$$
\frac{\partial \overline{g} (\omega, Q^2)}{\partial \log Q^2} =
\gamma_{gg} (\omega, \alpha_S) \: \overline{g} (\omega, Q^2)
$$
\noindent where the anomalous dimension
$$
\gamma_{gg} \equiv \int_0^1 dx \: x^\omega \: P_{gg} (x,
\alpha_S) \approx \frac{\overline{\alpha}_S}{\omega}
$$
\noindent if we use the small $x$ approximation:  $P_{gg} \approx
\overline{\alpha}_S/x$ with $\overline{\alpha}_S \equiv 3
\alpha_S/\pi$.  This is the double leading logarithm (DLL)
approximation, see Fig.\ 3.  The terms that are included in full
leading (and next-to-leading) order GLAP evolution are shown
connected by horizontal dotted lines.  On the other hand the BFKL
equation resums a different subset of terms
$$
\gamma_{gg} \simeq \sum_{n = 1}^\infty \: A_n \left (
\frac{\alpha_S}{\omega} \right )^n \rightarrow \sum_{n = 1} \:
A_n \: \alpha_S \frac{(\alpha_S \log 1/x)^{n - 1}}{(n - 1)!},
$$
\noindent that is an $\omega^{-n}$ behaviour transforms into a
$(\log 1/x)^{n - 1}$ behaviour.  Only these leading order $\log
1/x$ terms are known for $\gamma_{gg}$.  Interestingly the
coefficients $A_2 = A_3 = A_5 = 0$.  Leading order for
$\gamma_{qg}$ corresponds to the sum of $\alpha_S
(\alpha_S/\omega)^n$ terms, and here the coefficients are known.
In fact the BFKL increase\footnote{The reduction of the BFKL
$k_T$-factorized form, (\ref{a4}), to the collinear form $F_2
\sim P_{qg} \otimes g$ is discussed in refs.\ \cite{ch,km}.} of
$F_2 (\sim P_{qg} \otimes g)$ with decreasing $x$ appears to be
more due to the resummation in $P_{qg}$ (and in the coefficient
functions) than that for $g$.

Several numerical studies incorporating various $\alpha_S^n
\omega^{- m}$ contributions have been undertaken
\cite{ekl,ehw,bf2,frt}.  The perturbative QCD effects are
unfortunately masked by the lack of knowledge of the
non-perturbative input.  In particular, the results are very
dependent on the form of the \lq starting' parton distributions.
GLAP evolution from a singular $x^{- \lambda}$ input with
$\lambda \gapproxeq 0.3$ appears to be perturbatively stable.
The situation for a \lq flat' input is much more confused.
Moreover care must be taken to avoid drawing definitive
conclusions from a model dependent analysis in which the gluon
and sea quark {\it input} behaviours are assumed to be strongly
linked.  The problem is that at small $x$ we observe one
structure function $F_2^{ep} (x, Q^2)$ and yet we need to freely
parametrize both $g (x, Q_0^2)$ and $S (x, Q_0^2)$.  A series of
global analyses of deep inelastic data, systematically including
more and more terms of Fig.\ 3, may be revealing.
\section{Unified CCFM equation}
The CCFM equation \cite{ccfm} embodies both the BFKL equation at
small $x$ and GLAP evolution at large $x$.  It is based on the
coherent radiation of gluons, which leads to an angular ordering
of gluon emissions.  Outside the ordered region there is
destructive interference between the emissions.  For simplicity
we concentrate on small $x$.  Then the differential probability
for emitting a gluon of momentum $q$ is of the form
\begin{equation}
dP \sim \overline{\alpha}_S \Delta_R \frac{dz}{z} \frac{d^2
q_T}{\pi q_T^2} \: \Theta \: (\theta - \theta^\prime)
\label{a11}
\end{equation}
\noindent where successive gluon emissions occur at larger and
larger angles.  $\Delta_R$ represents the virtual corrections
which screen the $1/z$ singularity.  We can use (\ref{a11}) to
obtain a recursion relation expressing the contribution of $n$
gluon emission in terms of that of $n - 1$.  On summing, we find
that the gluon distribution satisfies an equation
\begin{eqnarray}
f(x, k_T^2, Q^2) & & = f^0 + \overline{\alpha}_S \int_x^1
\frac{dz}{z} \int \frac{d^2 q}{\pi q^2} \Delta_R \: \times
\nonumber \\
& & \Theta (Q - zq) \: f \left (\frac{x}{z}, |\mbox{\boldmath
$k$}_T + \mbox{\boldmath $q$}|^2, q^2 \right )
\label{a12}
\end{eqnarray}
\noindent which we may call the CCFM equation \cite{ccfm}.  The
angular ordering introduces an additional scale (which turns out
to be the hard scale $Q$ of the probe), which is needed to
specify the maximum angle of gluon emission.

When we unfold $\Delta_R$, so that the real and virtual corrections
appear on equal footing, and then take the leading $\log (1/x)$ approximation
we find (\ref{a12}) reduces to the BFKL equation for a gluon distribution
$f$ which is independent of $Q^2$ (note that $\Theta (Q -zq) \rightarrow 1$).
On the other hand in the large $x$ region $\Delta_R \sim 1$ and
$\Theta (Q - zq) \rightarrow \Theta
(Q - q)$, which leads to GLAP transverse momentum ordering.  If
we replace $\overline{\alpha}_S/z$ by $P_{gg}$ we see that
(\ref{a12}) becomes the integral form of the GLAP equation.

Explicit solutions $f (x, k_T^2, Q^2)$ of the CCFM equation have
recently been obtained in the small $x$ region \cite{kms}.  As
anticipated, the CCFM equation generates a gluon with (i) a
singular $x^{- \lambda}$ behaviour, with $\lambda \simeq 0.5$,
(ii) a $k_T$ distribution which broadens as $x$ decreases and
(iii) a suppression at low $Q^2$.  Fig.\ 4 compares the effective
$\lambda$ of the integrated gluon, that is
$$
xg (x, Q^2) \equiv \int^{Q^2} \: \frac{dk_T^2}{k_T^2} \: f (x,
k_T^2, Q^2) \: \sim \: x^{- \lambda},
$$
\ffig{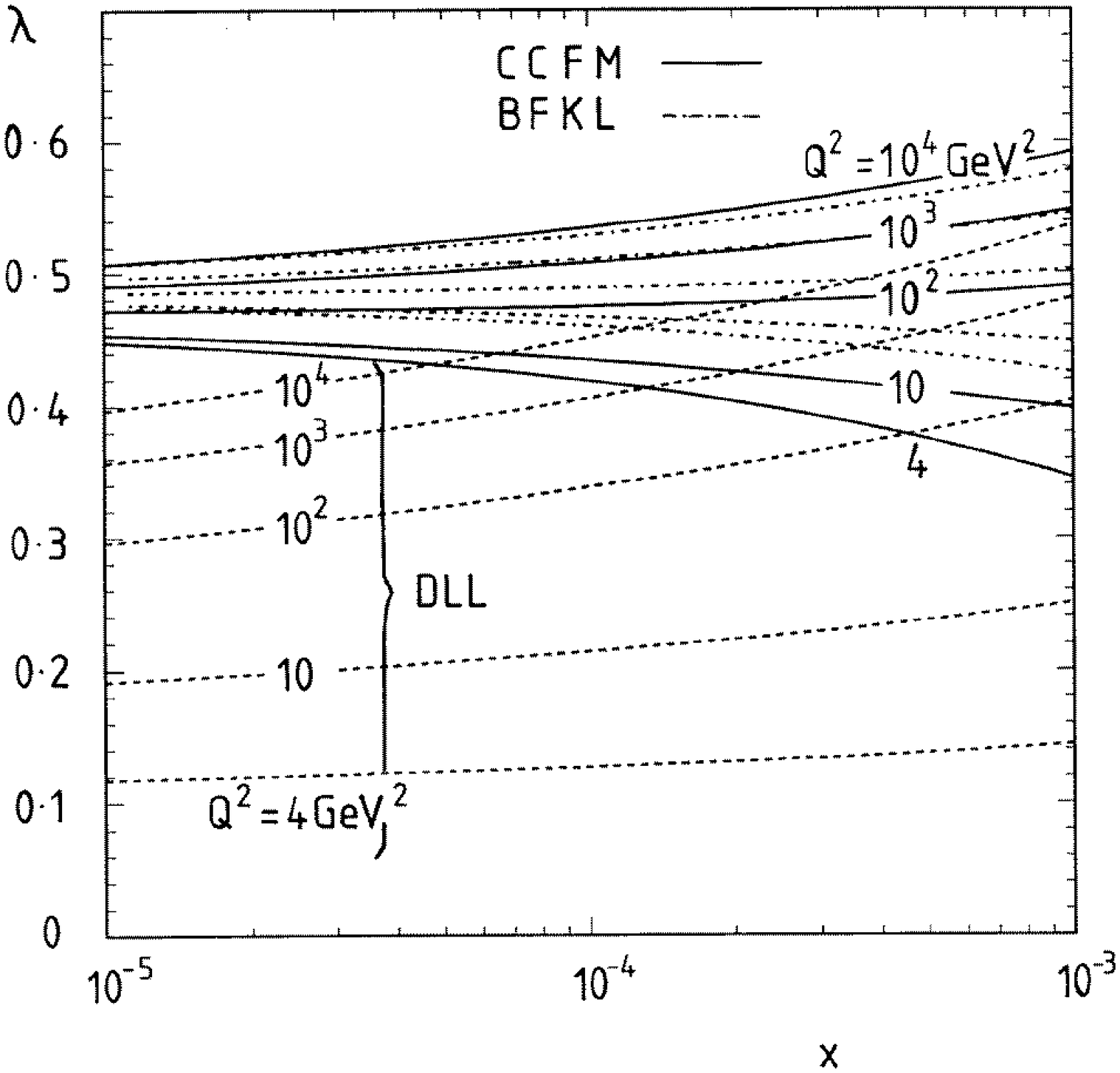}{60mm}{\em The effective values of $\lambda$,
defined by $xg = Ax^{- \lambda}$, obtained from the CCFM, BFKL
equations and the conventional DLL approximation, for different
values of $Q^2$.  The figure is taken from ref.\
[17].}{fig4}

\noindent obtained from the CCFM solution, with that from the
BFKL and DLL solutions.  Both the CCFM and BFKL values converge
to $\lambda \simeq 0.5$ at small $x$ independent of $Q^2$, in
contrast to DLL.  In addition we see the angular ordering,
embodied in the CCFM equation, leads to a suppression at low
$Q^2$.

\begin{center}
{\large\bf Acknowledgements}
\end{center}
I thank Jan Kwiecinski, Dick Roberts, James Stirling and Peter
Sutton for very enjoyable collaborations on the topic of this
review.
%

\end{document}